\newcommand{\jcapformat}{} 

\ifdefined\jcapformat
\documentclass[11pt,a4paper,pdflatex]{article}
\usepackage{jcappub}
\usepackage{float}
\else
\documentclass[usenatbib, twocolumn, nofootinbib, reprint,prd]{revtex4-1}
\usepackage{color}
\fi

\usepackage{newtxtext,newtxmath}
\usepackage[normalem]{ulem}
\usepackage{multirow}
\usepackage{booktabs}
\usepackage{amsmath,amsfonts,bm,hyperref}
\usepackage{graphicx}   
\usepackage[T1]{fontenc}
\usepackage{ae,aecompl}
\usepackage[normalem]{ulem}
\usepackage{dcolumn}
\usepackage{lineno}
\usepackage{seqsplit}
\newcolumntype{C}[1]{>{\centering\let\newline\\\arraybackslash\hspace{0pt}}m{#1}}

\newcommand{\lcdm}{$\Lambda$CDM}

\newcommand{\sqe}{sQE}
\newcommand{\iqe}{inpainted gradient}
\newcommand{\tszqe}{\tszfreemapnotation{} gradient}
\newcommand{\howmanyclusters}{5000}
\newcommand{\kappaest}{\hat{\bm{\kappa}}}
\newcommand{\kappaestrad}{\hat{\kappa}(\theta)}
\newcommand{\kapparad}{\kappa(\theta)}
\newcommand{\kappaonehalomz}{$\kappa(M,z)$}
\newcommand{\boxsize}{120^{\prime}}
\newcommand{\pixres}{0.^{\prime}5}
\newcommand{\innerradius}{4^{\prime}}
\newcommand{\outerradius}{30^{\prime}}

\newcommand{\mvir}{M_{200c}}
\newcommand{\mrecov}{M_{\rm lens}}
\newcommand{\massval}{2 \times 10^{14}\ \msol}
\newcommand{\zval}{0.7}

\newcommand{\msol}{\mbox{M}_{\odot}}
\newcommand{\munits}{\times 10^{14}\ \msol}

\newcommand{\snr}{$S/N$}
\newcommand{\bnhat}{\bm{\hat{\mbox{n}}}}
\newcommand{\bL}{\bm{\ell}}

\newcommand{\ukam}{${\rm \mu K^{\prime}}$}

\newcommand{\radbinsize}{1^{\prime}}
\newcommand{\totalradbins}{10}
\newcommand{\tszfreemapnotation}{tSZ-free}
\newcommand{\ccb}{{\it cluster convergence bias}}
\newcommand{\ccbnoitalics}{cluster convergence bias}

\newcommand{\planck}{{\it Planck}}
\newcommand{\sz}{Sunyaev-Zel{'}dovich}

\ifdefined\jcapformat
\newcommand{\apj}{The Astrophysical Journal}
\newcommand{\prd}{Physical Review D.}
\fi

\newcommand{\physrep}{Phys. Rept.}
\newcommand{\jcap}{JCAP}

\newcommand{\aap}{A\&A}
\newcommand{\araa}{ARA\&A}

\newcommand\mnras{MNRAS}

\defcitealias{raghunathan17a}{R17}
\defcitealias{raghunathan18}{R18}
\defcitealias{baxter18}{B18}
\defcitealias{sehgal10}{S10}
\defcitealias{mcclintock18}{M18}

\begin{document}

\title{An Inpainting Approach to Tackle the Kinematic and Thermal SZ Induced Biases in CMB-Cluster Lensing Estimators}

\newcommand{\authora}{Srinivasan Raghunathan}
\newcommand{\authorb}{Gilbert P. Holder}
\newcommand{\authorc}{James G. Bartlett}
\newcommand{\authord}{SanjayKumar Patil}
\newcommand{\authore}{Christian L. Reichardt}
\newcommand{\authorf}{Nathan Whitehorn}

\newcommand{\affSR}{Department of Physics and Astronomy, University of California, Los Angeles, CA, USA 90095}
\newcommand{\affGH}{Department of Astronomy and Department of Physics, University of Illinois, 1002 West Green St., Urbana, IL, USA 61801}
\newcommand{\affCR}{School of Physics, University of Melbourne, Parkville, VIC 3010, Australia}
\newcommand{\affSP}{School of Physics, University of Melbourne, Parkville, VIC 3010, Australia}
\newcommand{\affJBa}{APC, AstroParticule et Cosmologie, Universit\'e Paris
Diderot, CNRS/IN2P3, CEA/lrfu, Observatoire de Paris, Sorbonne Paris Cit\'e, 10, rue Alice Domon et L\'eonie Duquet, Paris
Cedex 13, France}
\newcommand{\affJBb}{Jet Propulsion Laboratory, California Institute of Technology, 4800 Oak Grove Drive, Pasadena, CA, USA 91109}
\newcommand{\affNW}{Department of Physics and Astronomy, University of California, Los Angeles, CA, USA 90095}

\ifdefined\jcapformat
\author[a]{\authora}
\author[b]{\authorb}
\author[c, d]{\authorc}
\author[e]{\authord}
\author[e]{\authore}
\author[a]{\authorf}

\affiliation[a]{\affSR}
\affiliation[b]{\affGH}
\affiliation[c]{\affJBa}
\affiliation[d]{\affJBb}
\affiliation[e]{\affSP}
\emailAdd{sri@physics.ucla.edu}
\emailAdd{gholder@illinois.edu}
\emailAdd{bartlett@apc.univ-paris7.fr}
\emailAdd{s.patil2@student.unimelb.edu.au}
\emailAdd{christian.reichardt@unimelb.edu.au}
\emailAdd{nwhitehorn@physics.ucla.edu}
\else
\author{\authora}\email{sri@physics.ucla.edu}
\affiliation{\affSR}
\author{\authorb}\affiliation{\affGH}\email{gholder@illinois.edu}
\author{\authorc}\affiliation{\affJBa, \affJBb}\email{bartlett@apc.univ-paris7.fr}
\author{\authore}\affiliation{\affCR}\email{christian.reichardt@unimelb.edu.au}
\author{\authorf}\affiliation{\affNW}\email{nwhitehorn@physics.ucla.edu}
\fi


\date{Accepted XXX. Received YYY; in original form ZZZ}

\newcommand{\abstracttext}
{A galaxy cluster's own \sz{} (SZ) signal is known to be a major contaminant when reconstructing the cluster's underlying lensing potential using cosmic microwave background (CMB) temperature maps. 
In this work, we develop a modified quadratic estimator (QE) that is designed to mitigate the lensing biases due to the kinematic and thermal SZ effects. 
The idea behind the approach is to use inpainting technique to eliminate the cluster's own emission from the large-scale CMB gradient map. 
In this inpainted gradient map, we fill the pixel values at the cluster location based on the information from surrounding regions using a constrained Gaussian realization. 
We show that the noise induced due to inpainting process is small compared to other noise sources for upcoming surveys and has negligible impact on the final lensing signal-to-noise. 
Without any foreground cleaning, we find  a stacked mass uncertainty of 6.5\% for the CMB-S4 experiment on a cluster sample containing \howmanyclusters{} clusters with $\mvir = 2 \munits$ at $z$ = 0.7. 
In addition to the SZ-induced lensing biases, we also quantify the low mass bias arising due to the contamination of the CMB gradient by the cluster convergence. 
For the fiducial cluster sample considered in this work, we find that this bias is negligible compared to the statistical uncertainties for both the standard and the modified QE even when modes up to $\sim 2700$ are used for the gradient estimation. 
With more gradient modes, we demonstrate that the sensitivity can be increased by 14\% compared to the fiducial result quoted above using gradient modes up to $2000$.
}

\ifdefined\jcapformat
\abstract\abstracttext
\else
\begin{abstract}
\abstracttext
\end{abstract}
\fi

\maketitle

\section{Introduction}
\label{sec_intro}
Galaxy clusters are remarkable cosmological probes,
with the caveat that their masses must be accurately measured. 
In general, cluster masses are not directly measured but estimated from mass-observable scaling relations with observed quantities such as the \sz (SZ) fluxes for microwave surveys \citep{bleem15, hilton18}; galaxy richnesses and galaxy velocities for optical surveys \citep{rozo10, saro13, ruel14}; or temperatures of the intra-cluster medium and luminosities for X-ray \citep{allen08, mantz10b} measurements. 
The mass precision of these methods has been steadily progressing although systematic uncertainties that dominate the cluster mass budget remain around that complex astrophysics that relate the survey observables to the underlying dark matter.
Once the mass is correctly measured, the abundance of galaxy clusters as a function of redshift and mass can reveal a great deal of information about the cosmological parameters that control the geometry and growth of structures in the Universe \citep[for a review, see][]{allen11}. 

Weak gravitational lensing can provide an unbiased probe of the matter distribution in a galaxy cluster and, arguably yields the best mass estimate. 
Consequently, several efforts have been made in the last decade to calibrate the mass-observable scaling relations mentioned above using lensing measurements \citep[recently,][]{mcclintock18, miyatake18, dietrich19}.
Weak-lensing measurements can be performed using two background light sources: galaxies or the cosmic microwave background (CMB).
While galaxy weak-lensing measurements provide excellent mass constraints for clusters below $z \le 1$, CMB-cluster lensing is more effective for distant galaxy clusters. 
Galaxy weak lensing struggles at these high redshifts due to the difficulty in imaging enough background (i.e. even higher redshift) galaxies at sufficient signal-to-noise (\snr). 
The CMB, in contrast, originates at $z = 1100$ and backlights all galaxy clusters at any redshift. 
Thus, CMB-cluster lensing is expected to yield the best mass measurements for the thousands of distant galaxy clusters \citep{raghunathan17a} expected to be detected by next generation CMB surveys \citep{cmbs4-sb1, SO18}.
Unfortunately though, the lensing \snr{} for a single cluster is low and one must rely on the stacked mass measurement to achieve a reasonable \snr{} for CMB-cluster lensing.

Lensing of the CMB by a galaxy cluster produces a unique dipole pattern that is oriented along the direction of the background CMB gradient at the cluster location.
Several estimators have been proposed in the literature \citep{seljak00, dodelson04, holder04, maturi05, lewis06, hu07, yoo08, melin15, horowitz19} to extract the lensing signal and the first detections were made by the Atacama Cosmology Telescope (ACT) \citep{madhavacheril15}, South Pole Telescope (SPT) \citep{baxter15}, and \planck{} \citep{placksz15} experiments. 
The measurements have flourished since then and have been used to: constrain the hydrostatic mass bias of clusters \citep{placksz15, hurier18, zubeldia19}; constrain the optical richness-mass ($\lambda-M$) relation of galaxy clusters \citep{geach17, baxter18, raghunathan18}; and estimate masses of halos hosting high redshift quasars \citep{geach19}.

Lensing measurements using CMB temperature data, however, are highly susceptible to foregrounds, the cluster \sz{} (SZ) signals in particular \citep{raghunathan17a}, and previous works have reduced the SZ-biases at the cost of the lensing \snr{} \citep{baxter15, baxter18}. 
In the standard quadratic estimator (\sqe), the lensing reconstruction is performed using two maps: one to measure the large-scale CMB gradient and the other map to measure the small-scale anisotropies \citep{hu07}. 
This estimator leverages the fact that gravitational lensing correlates different angular scales that are  uncorrelated in the primordial CMB anisotropy. 
The real world scenario is complicated by the cluster's own emission as such foreground signals are also correlated across different angular scales.  
But given that the lensing signal is oriented along the direction of the background CMB gradient while the foregrounds have no such dependence, it must be possible to decouple the foregrounds from biasing the lensing results.

The most important source of bias is the cluster thermal SZ (tSZ) signal. 
The tSZ effect is caused because of the interaction between CMB photons and the hot electrons inside the clusters and has a unique frequency dependance \citep{sunyaev70}. 
Recently, \citet{madhavacheril18} and \citet[][hereafter \citetalias{raghunathan18}]{raghunathan18} used this frequency dependance and modified the quadratic estimator to  estimate the background CMB gradient using a tSZ-free map. 
This \tszqe{} QE  eliminates the tSZ-induced correlation between the two maps, and thus removes the tSZ bias. 

There is also a second SZ effect due to the doppler shifting of CMB photons by the 
peculiar velocity of the cluster \citep{sunyaev80b}, known as the kinematic SZ (kSZ) effect. 
While smaller on average than the tSZ effect, the kSZ signal has been observed by several experiments \citep{hand12, mroczkowski13, sayers13, hill16, schaan16, soergel16, adam17, debernardis17}.
Unfortunately, the kSZ signal can not be removed in the same fashion as the tSZ effect because the kSZ effect has the same blackbody spectrum as the CMB. 
As a result, the  \tszqe{} QE mentioned above still suffers from a bias due to the kSZ effect. 
The level of kSZ bias is much smaller than the statistical errors for current measurements \citep{raghunathan18} but is expected to cause problems for cluster lensing measurements with future low noise CMB datasets \citep{raghunathan17a}. 

Other cluster emission can also bias the CMB-cluster lensing mass measurements. 
After the SZ effects, the most significant of these are the thermal dust and synchrotron emission from the cluster's member galaxies. 
Like the tSZ effect, one might depend on a linear combination of observing frequencies to remove these signals -- at the cost of reducing the \snr. 

Since the SZ effect and other foregrounds are largely unpolarized \citep{hall14, yasini16}, working with polarization based estimators is another effective way to handle these foreground biases.
Polarization estimators have the advantage of foreground simplicity, but are handicapped by the fact that,  for CMB experiments over the next decade, the temperature measurements have much higher potential \snr. 
An alternative approach is to combine the temperature and polarization measurements by simply replacing the large-scale CMB gradient map with the polarization $E/B$-mode maps. 
However, this approach does not yield the same \snr{} as temperature-only measurements as the correlation between the primordial CMB temperature and polarization fields is $\lesssim 20\%$.

In this work, we design a modified QE that is robust against bias from all types of cluster emission. 
We focus on performance with respect to the kSZ and tSZ effect biases, but the scheme should also mitigate dust and radio emission. 
The idea is to use an inpainted map for the large-scale CMB gradient estimation. 
In this approach, we replace the pixel values of the CMB gradient map near the cluster location using information from the surrounding regions. 
We show that the estimator yields unbiased lensing mass estimates on simulations. 
While, in principle, the inpainting process can introduce additional variance, we show that is minimal and has negligible impact on the lensing \snr{} compared to the \sqe.
Inpainting eliminates the spatially-localized foregrounds and distortions, while maintaining the harmonic structure of background CMB. 
With no cluster SZ signals in the inpainted gradient map, there will no longer be foreground-induced correlations between the two maps used in the QE and as a result no bias in the lensing reconstruction. 
We also study the bias due to SZ signals beyond the typical cluster size chosen for inpainting that originate from the haloes correlated to the cluster and find them to be negligible. 
The injected SZ foregrounds for these tests are extracted from N-body simulations \citep{sehgal10} to capture all real-world effects.
Finally, an added advantage of inpainting is that it also allows us to estimate the unlensed CMB gradient without being influenced by the cluster's own convergence field. 
\citet{hu07} noted that the magnification created by cluster lensing led to an underestimation of the background CMB gradient at the cluster location. 
We quantify this mass bias in the \sqe{} and compare it to the \iqe{} estimator.

The paper is organized as follows. 
The standard quadratic lensing estimator, its modifications, and a brief description of the inpainting technique are described in \S\ref{sec_lensing_estimator}. 
This is followed by an outline of simulations used to validate the cluster lensing pipeline in \S\ref{sec_sims_overview}. 
In \S\ref{sec_baseline}, we compare our estimator to the \sqe{} and \tszqe{} estimators, and then quantify the effect of cluster convergence bias.
The tests to check robustness of the estimator to foreground induced lensing biases are in \S\ref{sec_sz}.
We discuss the potential of the estimator with future CMB datasets and conclude in \S\ref{sec_conclusion}. 
Throughout this work adopt the \lcdm{} cosmology obtained from \planck{} chain that combines the \planck{} 2015
temperature, polarization, lensing power spectra with BAO, H0, and SNe data (TT,TE,EE+lowP+lensing+ext
in Table 4 of \citet{planck15-13}). 

\section{Methods}
\label{sec_methods}
We introduce the standard lensing quadratic estimator and then present the inpainting technique. 
This is followed by an overview of the simulations used to validate the lensing pipeline.

\subsection{Lensing estimator}
\label{sec_lensing_estimator}
\subsubsection{The quadratic estimator for CMB-cluster lensing}

Gravitational lensing of the CMB remaps the unlensed CMB temperature or polarization fields $\tilde{X}(\bnhat)$ ($X \in [T, E, B]$) to new positions based on the deflection angle $\alpha(\bnhat) = \nabla \phi$ corresponding to the underlying lensing potential $\phi$. 
The total power is conserved by gravitational lensing. 
Thus, the observed CMB field $X$  at a given position can be written as \citep{zaldarriaga99}
\begin{eqnarray}
X(\bnhat) = \tilde{X}[\bnhat + \nabla \phi(\bnhat))] \sim \tilde{X}(\bnhat) + \nabla\tilde{X}(\bnhat) \cdot \nabla\phi(\bnhat),
\label{eq_lensing_taylor}
\end{eqnarray} where we have dropped higher order terms in the Taylor expansion  to obtain the expression on the right. 

It is convenient to transform the CMB and deflection angle fields to Fourier space, $\tilde{X}_{\ell}$ and $\alpha_{\ell}$, where we have used the subscript $\ell$ to indicate the Fourier transformations. 
In Fourier space, lensing convolves the unlensed CMB field $\tilde{X}_{\ell}$ and the underlying deflection angle field $\alpha_{\ell}$. 
Thus, lensing correlates the otherwise independent modes.
The basic idea behind the QE is to use  these lensing induced correlations to reconstruct the underlying lensing convergence signal $\kappa = -\nabla^{2}\phi$. 

The QE looks at the product of two maps, $G$ and $L$,  to reconstruct the lensing signal \citep{hu07}: 
\begin{equation}
\hat{\kappa}_{\bL} = -A_{\ell} \int d^{2}\bnhat\ e^{-i\bnhat\cdot\bL}\ {\rm Re} \left\{ \nabla \cdot \left[ G(\bnhat) L^{*}(\bnhat)\right]\right\}. 
\label{eq_QE_kappa}
\end{equation}
The two maps, which would be independent in the absence of lensing,  are both derived from the CMB temperature map $T$ according to:
\begin{eqnarray}
G_{\bL} &=& i\bL{} \,W_{\ell}^{G}\, T^{G}_{\bL},\\
L_{\bL} &=& W_{\ell}^{L} \,T^{L}_{\bL}. 
\label{eq_QE_filtered_gradient_lensing_maps}
\end{eqnarray}
Here the first map $G_{\bL}$ corresponds to a large-scale gradient map while $L_{\bL}$ is a CMB temperature map weighted optimally to extract the small-scale lensing signal.
The notational choice here to distinguish $T^{G}$ and  $T^{L}$ prefigures the modified versions of the QE; in the original estimator the same temperature map is used for both legs, $T  \equiv T^{G} = T^{L}$. 
The optimal linear filters $W_{\ell}^{G}$ and $W_{\ell}^{L}$ are designed  to maximize the lensing \snr{}  \citep{hu07}:
\begin{eqnarray}
W_{\ell}^{G} &=&   \left\{
\begin{array}{l l}
C^{\rm unl}_{\ell} (C_{\ell} + N_{\ell}^{G})^{-1}&, ~\ell \le \ell_{\rm G}\\\notag
0&, ~{\rm otherwise}
\end{array}\right.\\\notag\\
W_{\ell}^{L} &= & (C_{\ell} + N_{\ell}^{L})^{-1}.
\label{eq_QE_filters}
\end{eqnarray} 
In the above equation, $(C_{\ell}^{{\rm unl}})C_{\ell}$ represent the (un)lensed CMB temperature power spectra. 
In this work, we calculate these spectra using 
 the Code for Anisotropies in the Microwave Background (\texttt{CAMB}\footnote{\url{https://camb.info/}}]) software \citep{lewis00}.
$N_{\ell}$ is the beam-deconvolved noise power spectrum for a given map. 

Note that in Eq.~(\ref{eq_QE_filters}), the weights for the gradient modes above $\ell_{G}$ are set to zero. 
As pointed out by \citet{hu07}, the CMB gradient at the cluster location is under-estimated due to the magnification of the CMB by the underlying cluster convergence and this leads to bias in the reconstructed cluster mass. 
Henceforth, we will refer to this effect as the \ccb.
Since scales larger than a typical cluster scale (few arcminutes) are relatively unaffected by the convergence signal, the bias can be mitigated by simply removing the more contaminated small-scale modes in the gradient estimation.
Subsequently, several works have set $\ell_{G}=2000$ \citep{madhavacheril15, horowitz19, raghunathan18} in their analyses following \citet{hu07}. 
This heuristic choice of $\ell_{G} = 2000$ comes from the fact that the small-scale power of the unlensed CMB is exponentially damped \citep{silk68} and excluding gradient modes $\ell \ge 2000$ does not significantly degrade the lensing \snr{} \citep[also see Fig. 1 of][]{hu07}. 
We return to this discussion in \S~\ref{sec_ngrad} and demonstrate that inpainting allows more modes to be included (and thus adds sensitivity) without being affected by the \ccbnoitalics.


\subsubsection{Modification to the QE}
\label{sec_inpainting}

As mentioned earlier, the bias due to the cluster tSZ signal can be mitigated by estimating the background CMB gradient from a \tszfreemapnotation{} CMB map.
For example, \citetalias{raghunathan18} combined the 95 and 150 GHz channels from the SPTpol experiment to clean the tSZ signal from the gradient map.
While the \tszfreemapnotation{} gradient approach works, in principle, to eliminate the bias due to cluster tSZ signal, it is not suitable to remove the lensing induced bias due to cluster kSZ signal as, unlike tSZ, kSZ has no average frequency dependence. 

In this work, we follow a different approach. 
We replace the \tszfreemapnotation{} gradient map with a \iqe{} map in which we fill the pixel values at the cluster location with the information from the surrounding region. 
This inpainted gradient approach has the following two key merits compared to a \tszqe{}. 
First, the approach can help in mitigating the kSZ-induced lensing bias; 
and second, the noise and other frequency dependent foregrounds in the gradient map are not enhanced. 



Based on \citet{benoitlevy13} we define two regions at distances $R \le R_{1}$ and $R_{1} < R \le R_{2}$ from the cluster center.
We fill the CMB temperature values $T_{1}$ in region $R_{1}$ based on the temperature $T_{2}$ in $R_{2}$ using a constrained Gaussian realization as 
\begin{equation}
\hat{T}_{1} = \tilde{T}_{1} + {\bf \hat{C}}_{12}  {\bf \hat{C}}_{22}^{-1} (T_{2} - \tilde{T}_{2})
\end{equation} where $\tilde{T}_{1}, \tilde{T}_{2}$ are the temperature values in the two regions from a random Gaussian realization of the CMB and we use hats to represent the estimated quantities.
The above operation is possible since the primordial CMB exhibits coherence over degree scales and the covariance matrix ${\bf \hat{C}}_{XY}$ of the CMB fields between two regions $X, Y$ can be calculated using simulations as
\begin{eqnarray}
 {\bf \hat{C}}_{XY} =  {\bf \hat{C}}_{YX} = \frac{1}{n-1}\sum\limits_{i = 0}^{n} \left( {\bf G}_{i}- \left< {\bf G} \right>\right) \left( {\bf G}_{i} - \left< {\bf G} \right>\right)^{T}
\label{eq_cmb_cov_T1T2}
\end{eqnarray} where the index $i$ runs over the number of simulations and
\mbox{${\bf G} = \left(
\begin{array}{c c} {\bf G}_{X}\\ {\bf G}_{Y}\end{array}
\right)$} 
is the concatenated vector of a Gaussian realization of the large-scale structure (LSS) lensed CMB map in the two regions. The details about our simulations can be found in \citet{raghunathan17a} and are discussed briefly in the following section \S~\ref{sec_sims_overview}. 

The effectiveness of the estimator depends on how close the inpainted values $\hat{T}_{1}$ are to the true background CMB gradient $T_{1}$. 
Of course, inpainting introduces extra noise and, as expected, the difference between true and inpainted values is higher when the input map is noisier. 
In addition, the process of inpainting introduces mode mixing and, as a result, even though we are only interested in reconstructing the large-scale gradient modes, it is not trivial to avoid the contamination from the noisy small-scale modes. 
Consequently, we low-pass filter (LPF) the input CMB map $T$ to remove modes $\ell > \ell_{\rm LPF}$ before performing the inpainting process. 

\begin{figure*}
\centering
\ifdefined\jcapformat
\includegraphics[width=1\textwidth, keepaspectratio]{figs/inpainting_example.eps}
\else
\includegraphics[width=0.47\textwidth, keepaspectratio]{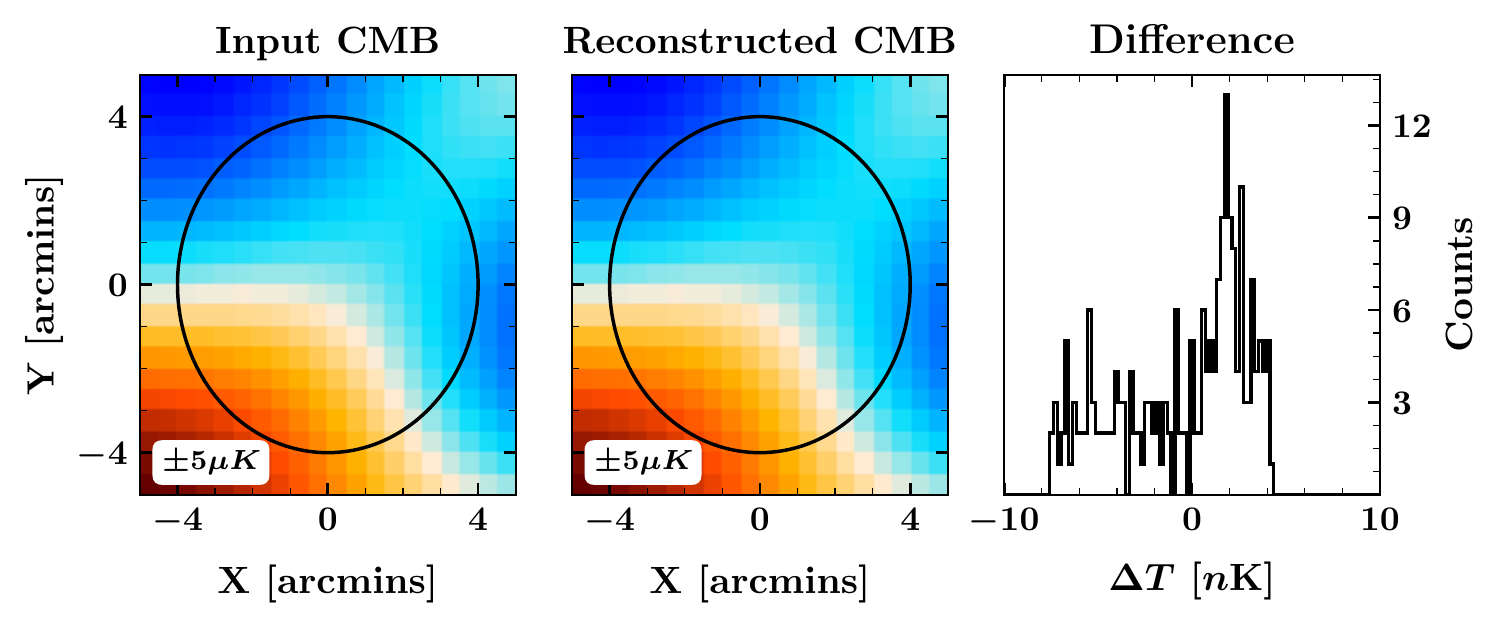}
\fi
\caption{Illustration of the reconstructed CMB temperature using the inpainting technique. The first two panels are $10^{\prime} \times 10^{\prime}$ cutouts of the input and the reconstructed signals. 
They have been LPF at $\ell_{\rm LPF, R1}$ and include signals from CMB, foregrounds, and noise.
The black circle is the ($R1 = \innerradius$) inpainted region. 
The right panel shows the pixel differences between the two within the inpainted region in nano-Kelvin units and is much smaller than the noise in the input CMB map in the left panel.
} 
\label{fig_inpainting_example}
\end{figure*}

We note that the size of the inpainting radius $R1$, in principle, depends on the angular size of the cluster and must be chosen dynamically based on the cluster mass and redshift in case of real data. 
Here, we conservatively set $R1 = \innerradius$ to encompass the clusters from a wide range of masses and redshifts. 
As an example, for the clusters considered in the current work $R1 \sim \theta_{200c}$ where $\theta_{200c} = R_{200c}/D_{A}(z)$ with $D_{A}$ being the angular diameter distance to the cluster redshift $z$. 
For massive nearby clusters, however, $R1$ must be higher as we demonstrate in later sections. 
For the LPF, we use $\ell_{\rm LPF} \equiv \ell_{\rm LPF, R1} = 2700$ which corresponds to the size of the inpainting radius $R1$. 
The choice of $ \ell_{\rm LPF, R1}$ is well motivated without much \snr{} penalty given that: (a) no information can be recovered from the inpainted map below the scale corresponding to $R1$; and (b) the small-scale primordial CMB is exponentially damped \citep{silk68}. 
The only drawback of the LPF is the undesired ringing effects in the presence of a bright SZ signal at the cluster location. 
In \S\ref{sec_sz}, we describe additional filtering steps in the input CMB map $T$ to attenuate these filtering artifacts and cleanly reconstruct the cluster convergence.
The outer region $R2$ is set to $\outerradius$ and we perform 50,000 simulations to estimate the covariance matrix. 
We have confirmed that increasing R2 has negligible impact on our results. 
An illustration of the inpainting process is given in Fig.~\ref{fig_inpainting_example} where the first two panels are the $10^{\prime} \times 10^{\prime}$ cutouts from the original and reconstructed CMB maps around the desired region chosen for inpainting. 
Both the panels include signals from CMB, and variance from noise/foregrounds and have been LPF at $\ell_{\rm LPF, R1}$. 
The histogram in the right is the difference between the two panels within $R1$ (black circle). 
The residual shown in nano-Kelvin units is negligible compared to the noise in the input CMB map.

\subsection{Simulation set-up}
\label{sec_sims_overview} 
Following the previous work \citep{raghunathan17a}, we create Gaussian realizations of the CMB temperature map using the LSS lensed power spectra $C_{\ell}^{TT}$ generated using \texttt{CAMB} \citep{lewis00} software for the fiducial \planck\ 2015 cosmology. 
In these simulations, we neglect the small non-Gaussianities arising due to the lensing from intervening LSS.
The simulations are done in the flat-sky approximation and are $\boxsize\ \times\ \boxsize$ wide with a pixel resolution of $\pixres$. 
The simulations are then lensed using galaxy cluster convergence profiles described below. 
We add Gaussian realizations of the foregrounds not associated to the galaxy cluster using the measurements made by the SPT-SZ experiment \citep{george15}. 
The foregrounds include dusty and radio galaxy emissions, and the SZ signals from the unresolved haloes in the SPT-SZ 90 and 150 GHz channels.
In \S\ref{sec_sz}, we also include the cluster SZ signals to quantify the level of bias induced in the reconstructed lensing signal. 
In all cases, the foregrounds remain unlensed by the galaxy cluster.
Next, we convolve the simulated maps with a Gaussian beam with $\theta_{\rm FWHM} = 1^{\prime}.4$ and add instrumental noise of $\Delta_{T} = 1.5$\ukam{} matching the expected configuration\footnote{We use the projected performance of the CMB-S4 150 GHz channel for the Large-area Survey 03 specified in the following link as on 13 March 2019: \url{https://cmb-s4.org/wiki/index.php/Survey_Performance_Expectations\#Large-area_Survey_Performance_Expectation_03}} of the wide area survey of the CMB-S4 \citep{cmbs4-sb1} experiment. 
For the \tszfreemapnotation{} maps, we set the noise to a slightly higher level $\Delta_{T} = 4.6$\ukam{} and use $\theta_{\rm FWHM} = 2^{\prime}$. 
This higher noise level is chosen to match the CMB-S4 \tszfreemapnotation{} map made from the combination of 90 and 150 GHz channels. 
Adding data from other available channels will tend to reduce the noise level slightly.

To model the galaxy cluster profile, we use the Navarro-Frenk-White (NFW) \citep{navarro96} formalism given in Eq.~(\ref{eq_nfw_density_with_delta_c})
\begin{eqnarray} 
\rho\left(r\right) & = & \frac{\rho_0}{\left(\frac{r}{R_{\rm s}}\right)\ \left(1+\frac{r}{R_{\rm s}}\right)^2},
\label{eq_nfw_density_with_delta_c}
\end{eqnarray} 
where $R_{\rm s}$ is the scale radius, $\rho_{0}$ is the central cluster density, and $c_{200}(M,z) = R_{200}/R_{\rm s}$ is the halo concentration parameter. 
To obtain the NFW convergence profile \kappaonehalomz{} for the NFW halo, we simply plug in the analytic expression given in Eq. (2.8) of \citet{bartelmann96}. 
We quote the cluster masses with respect to the radius $R_{200c}$, which encompasses the region within which the average mass density is 200 times the critical density of the Universe at the cluster redshift $z$. 
For simplicity, we fix the cluster masses and redshifts to: $\mvir = \massval$ and $z = \zval$ with $c_{200} = 3.12$ calculated using \citet{duffy08} formula.  


\begin{figure*}
\centering
\ifdefined\jcapformat
\includegraphics[width=1\textwidth, keepaspectratio]{figs/kappa_reconstructed_residual_example_simple_TF.eps}
\else
\includegraphics[width=1\textwidth, keepaspectratio]{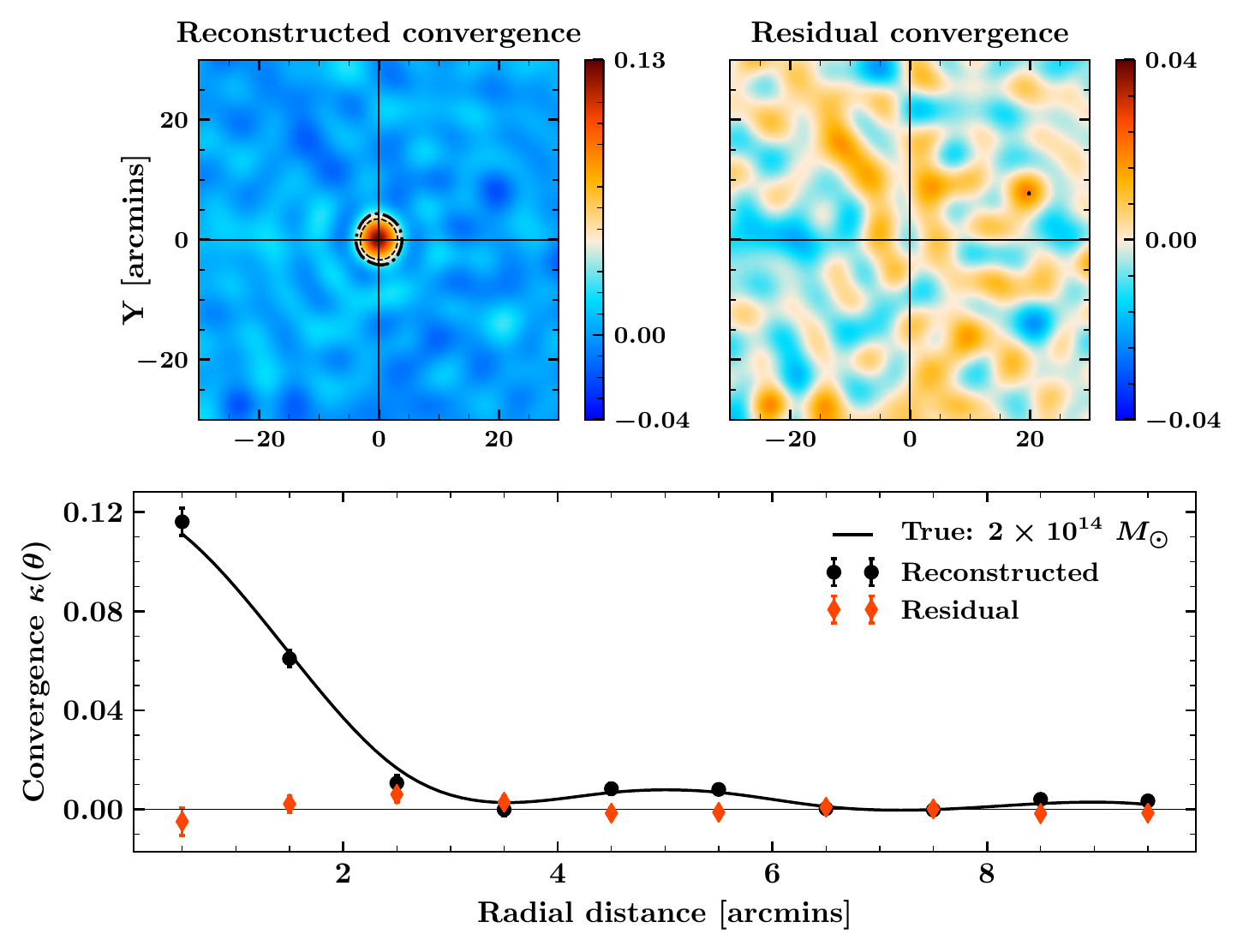}
\fi
\caption{
Top left panel shows the reconstructed stacked convergence maps of \howmanyclusters{} clusters using the \iqe{} estimator. 
The difference between the true cluster convergence and the left panel is shown in the top right panel. 
The radially-binned reconstructed (black circles), true (black solid curve), and residual (orange diamonds) convergence profiles are in the bottom panel. 
The residual signal is consistent with a null signal, probability to exceed (PTE) = 0.55, implying that the reconstructed convergence signal is unbiased.
The contour lines correspond to regions that are $\ge 3, 5\sigma$ from the background. 
}
\label{fig_QE_std_tszfree_innpainted_kappa_example}
\end{figure*}

\section{Results}
\label{sec_results} 

In \S~\ref{sec_baseline}, we compare the \iqe{} to the \sqe{} and \tszqe{} QE followed by a discussion about approaches to improve lensing \snr. 
Next, in \S~\ref{sec_sz}, we quantify the impact of the cluster SZ signals on the reconstructed lensing maps and how they can be mitigated using the \iqe{} estimator.

Throughout this section, we quote the mass constraints obtained by stacking the reconstructed lensing signal $\kappaest$ from \howmanyclusters{} galaxy clusters. 
The lensing masses quoted in the subsequent figures are the median values obtained from ten such simulations, each with \howmanyclusters{} stacked clusters. 
We bin the stacked signal into \totalradbins{} linearly-spaced radial bins $\kappaestrad$ with $\Delta\theta$ = $\radbinsize$. 
We compare the radially-binned $\kappaestrad$ to the fiducial NFW convergence \kappaonehalomz{} profiles binned radially into $\kapparad$ and obtain cluster mass constraints using
\ifdefined\jcapformat
\begin{equation}
-2\ln\mathcal{L (\hat{\kappa}(\theta)}| M) = 
\sum_{\theta = 0}^{\totalradbins^{\prime}}\left[\hat{\kappa}(\theta) - \kappa(\theta)\right] {\bf \hat{C}}^{-1} \left[\hat{\kappa}(\theta) - \kappa(\theta)\right]^{T}
\label{eq_QE_likelihood}
\end{equation}
\else
\begin{multline}
-2\ln\mathcal{L (\hat{\kappa}(\theta)}| M) = \\
\sum_{\theta = 0}^{\totalradbins^{\prime}}\left[\hat{\kappa}(\theta) - \kappa(\theta)\right] {\bf \hat{C}}^{-1} \left[\hat{\kappa}(\theta) - \kappa(\theta)\right]^{T}
\label{eq_QE_likelihood}
\end{multline}
\fi
To obtain the covariance matrix we use a jackknife re-sampling technique
\begin{equation}
{\bf \hat{C}} = \frac{N-1}{N} \sum\limits_{j=1}^{N = 1000} \left[\hat{\kappa}_{j}(\theta) - \left<\hat{\kappa}(\theta)\right>\right] \left[\hat{\kappa}_{j}(\theta) - \left<\hat{\kappa}(\theta)\right>\right]^{T},
\label{eq_JK_cov}
\end{equation}
where $\hat{\kappa}_{j}(\theta)$ is the azimuthally binned stacked convergence of all the clusters in the $j^{th}$ sub-sample and  $\left<\hat{\kappa}(\theta)\right>$ is the ensemble average of all the 1000 sub-samples. 
For parameter constraints, we sample the likelihood space and report the median mass values and $1\sigma$ uncertainties from the 16$^{th}$ and 84$^{th}$ percentiles.


\subsection{Results from idealized simulations}
\label{sec_baseline}
First, we discuss the mass constraints from all the three QE for idealized simulations with no cluster foreground signals and $\ell_{G} = 2000$. 
Next, we modify $\ell_{G}$ to include more gradient modes and quantify the \ccbnoitalics{} and the improvement in lensing \snr.
\subsubsection{Baseline setup with $\ell_{G} = 2000$}
\label{sec_baseline_lg}
For illustrative purposes, we show the reconstructed convergence map using the \iqe{} QE in Fig.~\ref{fig_QE_std_tszfree_innpainted_kappa_example}. 
The stacked convergence map is in the top left and the residual, difference between the true cluster convergence and the left panel, on the right.
In the bottom panel, we show radial profiles of the true (black curve), reconstructed (black circle), and residual (orange diamond) convergence signals. 
The PTE value of the residual signal is 0.55 (\mbox{$\chi^{2}_{\rm null}$ = 7.9} for 9 degrees of freedom), indicating that it is consistent with random fluctuations.

Next, we compare the mass constraints from three estimators in the top panel of Fig.~\ref{fig_mass_constraints}: \sqe{} (orange hexagon), \tszqe{} (red diamond), and \iqe{} (black circle).
All the three estimators return a cluster mass that is $\le 0.1\sigma$ from the input mass shown as black dash-dotted line. 
With the assumed experimental configuration and with no foreground cleaning for the 150 GHz channel, we note that the uncertainty in the stacked mass of our cluster sample is roughly $\Delta M/M \sim 6.5\%$. 
The fact that the \sqe{} and \iqe{} QE return similar constraints indicates that the noise from the inpainting process is negligible. 
Another subtle thing to note is that the \tszfreemapnotation{} gradient, created by combining multiple frequency channels has a higher noise ($\Delta_{T} = 4.^{\prime}5$\ukam) compared to the fiducial 150 GHz maps ($\Delta_{T} = 1.^{\prime}4$\ukam). 
This could result in a reduced \snr{} for the \tszqe{} QE compared to the others. 
At first glance, such a trend is not evident when restricting the gradient modes to $\ell_{G} = 2000$. 
We will come back to this discussion in following sections where we analyze the changes in the \snr{} upon modifying $\ell_{G}$.


\ifdefined\jcapformat
\begin{figure*}
\centering
\includegraphics[width=0.7\textwidth, keepaspectratio]{figs/mass_contraints_simple_TF.eps}
\else
\begin{figure}
\centering
\includegraphics[width=0.45\textwidth, keepaspectratio]{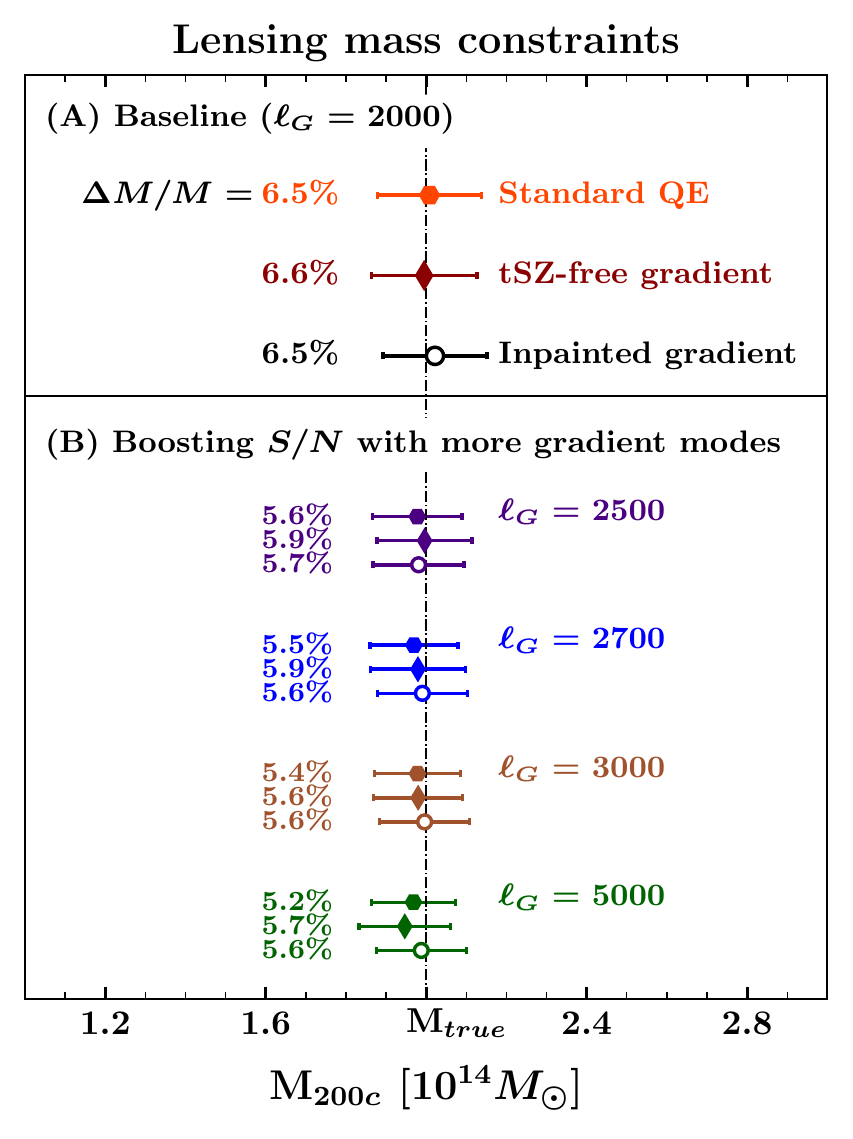}
\fi
\caption{
{\it Top panel:} The mass constraints obtained from the stacked convergence signal for the baseline case ($\ell_{G} = 2000$) with the \sqe{} (hexagon), \tszqe{} (diamond), and \iqe{} (circle) QE estimators. 
{\it Bottom panel:} The improvement in mass precision by including more modes for the gradient estimation. 
For the \iqe{} QE, the \snr{} saturates after $\ell_{G} = 2700$, which roughly corresponds to the scale of the inpainting radius $R1 = \innerradius$.}
\label{fig_mass_constraints}
\ifdefined\jcapformat
\end{figure*}
\else
\end{figure}
\fi

\subsubsection{Effect of cluster convergence bias}
\label{sec_mag_bias}
As mentioned in previous sections, the gradient estimation is limited to scales larger than $\ell \le \ell_{G}$ for QE in general. 
This LPF is introduced to mitigate the \ccbnoitalics{} arising from an underestimated gradient at the cluster location at the cost of a loss in sensitivity. 
In the baseline case in \S\ref{sec_baseline_lg}, we adopt $\ell_{G}= 2000$ which is reasonable for less massive clusters \citep{hu07}. 
But for massive low-redshift clusters which span a larger angular extent on the sky, $\ell_{G}= 2000$ may not be sufficient and can result in a mass bias of the order of few per cent \citep{hu07}. 
The goal of this section is to study the effect of the \ccbnoitalics{} for different cluster samples and understand if the sensitivity can be improved without being affected by the bias.

We start this test by investigating the recovered lensing mass from all three estimators for various values of $\ell_{G}$. 
This is shown in the bottom panel of Fig.~\ref{fig_mass_constraints}. 
For the fiducial setup, the effect of the magnification bias on modes $\ell < \ell_{G} (= 2000)$ is negligible.  
However, when the value of $\ell_{G}$ is increased (2000 through 5000), the recovered mass systematically shifts to lower values for both \sqe{} (hexagon) and \tszqe{} (diamond) points, which could be a hint of the \ccbnoitalics. 
The \iqe{} estimator (open circles), which is devoid of cluster convergence, is not affected and returns an unbiased mass for all values of $\ell_{G}$. 
While we impose $\ell_{G} > \ell_{\rm LPF, R1}$, we note that the size of the inpainting radius $R1$ acts as an ultimate LPF for the \iqe{} QE and no useful gradient information can be recovered from scales smaller than $\ell_{\rm LPF, R1} \ge 2700$. 

{\it Results for more massive cluster samples:} Since the level of bias for \sqe{} and \tszqe{} QE is much smaller than the statistical errors, the results are inconclusive for our fiducial cluster sample. 
So we extend our analysis to more massive clusters to exaggerate  the cluster convergence bias, and measure it in a computationally reasonable sample size. 
For this test, we use samples of clusters that are $4\times10^{14}~\msol$ (i.e.~ twice as massive as our fiducial sample) at redshifts $z = 0.7$ and $z=1.5$.  
The results are presented in Tab.~\ref{tab_mag_bias}. 
We chose two different radii $R1 = \innerradius$ and $5^{\prime}$ for the \iqe{} estimator and, to make a fair comparison, set $\ell_{G} = 2700$ and $2160$ corresponding to the scale of $R1$ in the \sqe. 
Our results suggest that the systematic bias due to cluster convergence is smaller than the statistical uncertainty for both the \sqe{} and the \iqe{} when $\ell_{G}$ is set to 2700 or smaller. 
When $\ell_{G} \ge 3000$ in \sqe, the bias increases and becomes comparable to the statistical errors. 
These results are consistent with the observations from Fig.~\ref{fig_mass_constraints} for the fiducial cluster sample. 
For even massive clusters with $\mvir = 8\ \munits$ at $z=0.7$, the bias is evident. 
The lensing inferred mass is shifted low by 7\% for both the \sqe{} ($\ell_{G} = 2700$) and \iqe{} ($R1 = \innerradius$) estimators. 
The bias vanishes when $R1$ is set to $8^{\prime}$ for \iqe{} or equivalently $\ell_{G} = 1350$ for \sqe{} with a significant ($\times\ 1.75$) hit in \snr{}. 

Based on the above results we conclude that it is safe to set $\ell_{G} \sim 2700$ when trying to reconstruct the convergence signal, of clusters with \mbox{$\mvir \lessapprox 4\times10^{14}~\msol$}, using next-generation datasets without worrying about convergence bias. 
The \iqe{} QE, however, does not have specific advantage over the \sqe{} with respect to the \ccbnoitalics.
The gain in the lensing \snr{} with a higher $\ell_{G}$ is discussed in the next section.

\begin{table}
\centering
\caption{Bias in the reconstructed lensing masses due to the cluster convergence signal. 
We consider two clusters samples with $\mvir = 4 \munits $ at redshifts $z=0.7$ and $z=1.5$.
}
\vspace*{2mm}
\footnotesize{
\ifdefined\jcapformat
\begin{tabular}{| l | C{1.6cm} |  C{2.cm}| C{2.6cm}|  C{2.cm}| C{2.6cm}|}
\else
\begin{tabular}{| C{1.8cm} | C{1.25cm}|  C{2.5cm}| C{2.5cm}| |  C{2.5cm}| C{2.5cm}| }
\fi
\hline
\multirow{2}{*}{Estimator} & \multirow{2}{*}{$\ell_{G}$} & \multicolumn{2}{c|}{$z=0.7$} & \multicolumn{2}{c|}{$z=1.5$}\\
\cline{3-6}
& &  Lensing mass [$10^{14}\ \msol$] & Bias \% (frac. of $\sigma$) & Lensing mass [$10^{14}\ \msol$] & Bias \% (frac. of $\sigma$) \\\hline
 & 2700 & $3.99 \pm 0.14$ & 0.2\% ($0.1\sigma$) & $3.96 \pm 0.13$ & 1.0\%  ($0.3\sigma$) \\
Inpainted & ($R1 = \innerradius$ ) & & & &\\
\cline{2-6}
gradient QE & 2160 & $4.01 \pm 0.15$ & 0.2\% ($0.1\sigma$) & $3.99 \pm 0.14$ & 0.2\% ($0.1\sigma$) \\
& ($R1 = 5^{\prime}$) & & & & \\\hline
\hline
\multirow{3}{*}{Standard QE} & 3000 & $3.93 \pm 0.13$ & 1.8\% ($0.5\sigma$) & $3.81 \pm 0.12$ & 4.8\%  ($1.5\sigma$)\\
\cline{2-6}
& 2700 & $4.04 \pm 0.13$ & 1.1\% ($0.3\sigma$) & $3.93 \pm 0.12$ & 1.8\% ($0.6\sigma$)\\
\cline{2-6}
& 2160 & $4.05 \pm 0.15$ & 1.2\% ($0.3\sigma$) & $4.03 \pm 0.14$ & 1.1\% ($0.2\sigma$)\\\hline
\end{tabular}
}
\label{tab_mag_bias}
\end{table}

\subsubsection{Boosting lensing {\it S/N} with more gradient modes}
\label{sec_ngrad}
Now that we have an estimate of the \ccbnoitalics, we could improve the \snr{} of the estimated convergence by including more modes in the gradient estimation. 
This improvement is shown in the bottom panel of Fig.~\ref{fig_mass_constraints}, where we show the mass constraints 
for different values of $\ell_{G}$ from all the three QE.
From the figure, it is evident that the \snr{} increases for all estimators when more gradient modes are included. 
The rate at which the \snr{} improves and the saturation point, however, differs.

When $\ell_{G}$ is set to 2500 (purple), we obtain 15\% better mass constraints for \sqe{}, 9\% for \tszqe{}, and 14\% for \iqe{} compared to the fiducial cases in the top panel with $\ell_{G} = 2000$. 
The reason for a significantly smaller improvement in \tszqe{} QE is attributed to the higher noise in \tszfreemapnotation{} map compared to the 150 GHz or the inpainted gradient map. 
The rate at which the \snr{} improves is only gradual for even higher $\ell_{G}$ due to the diffusion damping of the small-scale CMB \citep{silk68} (also see Fig. 1 of \citet{hu07}). 

There is no improvement in \snr{} for the \iqe{} when $\ell_{G} \ge 2700$, emphasising no useful gradient information can be recovered from scales smaller than the size of the inpainting radius $R1 = \innerradius$. 
As expected, the \snr{} saturates at a higher $\ell_{G}$ for both \sqe{} and \tszqe{} estimators. 
Since the \tszqe{} map is noisier than the 150 GHz map, the \tszqe{} QE attains saturation earlier than the \sqe.



\subsection{Effect of cluster correlated foregrounds}
\label{sec_sz}
\ifdefined\jcapformat
\begin{figure*}
\centering
\includegraphics[width=.7\textwidth,keepaspectratio]{figs/mass_contraints_SZbias_allestimators_simple_TF.eps}
\else
\begin{figure}
\centering
\includegraphics[width=0.45\textwidth, keepaspectratio]{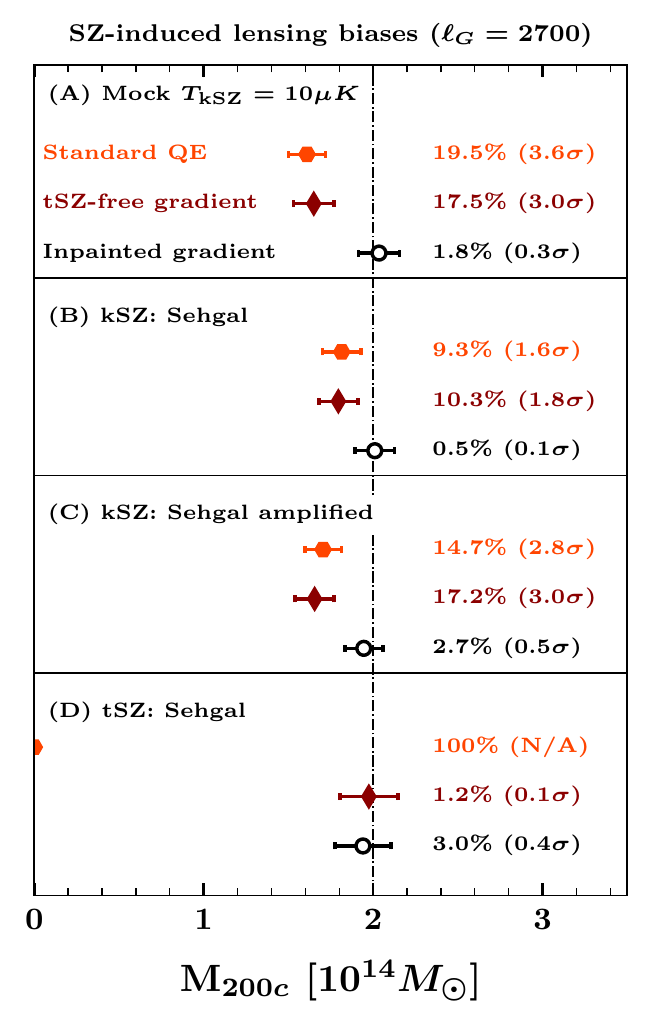}
\fi
\caption{
Same as Fig.~\ref{fig_mass_constraints}, but after including cluster SZ signals in the CMB maps. 
The data points for the standard, \tszqe{}, and \iqe{} are shown as orange hexagons, red diamonds, and black circles. 
The top three panels are for varying level of kSZ signals: top panel is for a mock kSZ with $T_{\rm kSZ} = 10 \mu K$;  second panel is using kSZ from \citetalias{sehgal10} simulations; third panel also uses \citetalias{sehgal10} simulations but after amplifying them using the signals corresponding to clusters with mass $4\munits$. 
In all cases, the \iqe{} provides an unbiased reconstruction of the cluster mass while the bias in the \sqe{} and \tszqe{} grow with the amplitude of the kSZ signal. 
The bottom panel is for tSZ from \citetalias{sehgal10} simulations. 
The \sqe{} is heavily biased due to cluster tSZ while the \tszqe{} and \iqe{} are unbiased.
}
\label{fig_mass_constraints_SZ_biases}
\ifdefined\jcapformat
\end{figure*}
\else
\end{figure}
\fi

Now we turn to the effect of cluster-correlated foregrounds: kSZ and tSZ signals from the galaxy cluster on the standard, \tszqe{}, and the \iqe{} QE. 
As discussed before, the cluster foregrounds introduce a bias in the lensing reconstruction  using QE because of the correlations between foregrounds present in the maps in the two legs of the estimator. 
As a result, we can expect both the cluster tSZ and kSZ to bias the standard QE. 
The \tszqe{} QE should only be sensitive to the bias from cluster kSZ which cannot be cleaned by combining data from multiple frequencies. 
Cluster tSZ should have no effect in this case since, by construction, \tszqe{} QE is free from tSZ in the first leg of the estimator. 
As we show below, the \iqe{} QE is insensitive to both cluster tSZ and kSZ, as they are removed using the inpainting technique in the first leg. 
We use $\ell_{G} = 2700$ for all the tests in this section but also show the reduction in foreground bias for the \sqe{} with $\ell_{G} = 2000$.

First, we inject simple mock kSZ signals where we simply alter the pixel temperatures by a certain amount. 
This is followed by realistic foreground signals from \citet[][hereafter \citetalias{sehgal10}]{sehgal10} simulations.
To this end, we extract half-arcminute-resolution \mbox{$\boxsize\ \times\ \boxsize$} cutouts from the 90 and 150 GHz kSZ and tSZ 
\citetalias{sehgal10} simulations around halos corresponding to the cluster sample used here. 
We also check the effect of enhanced foregrounds using SZ signals from more massive clusters in the \citetalias{sehgal10} simulations.
We scale the tSZ power down from the \citetalias{sehgal10} simulations by a factor of 1.75 to match the \citet{george15} measurements.
These foreground cutouts are added to our simulated galaxy-cluster-lensed CMB datasets and processed in the standard manner as discussed in \S\ref{sec_sims_overview}. 

Note that as mentioned in \S\ref{sec_inpainting}, we LPF the true CMB map at $\ell_{\rm LPF, R1} = 2700$ before the inpainting process to reduce the mode coupling introduced from noisy higher multipoles. 
This approach works well in the absence of cluster emission as shown in previous sections. 
When the cluster emissions are included, particularly tSZ, the LPF can introduce ringing, making the inpainting process problematic. 
There are several known ways of handling this such as, tapering the data with an apodization window or altering the filter to have a smoother roll-off. 
In this work, we simply remove an estimate of the SZ signal from the gradient map before the LPF step. 
We explain this in the subsequent sections below. 
No such 
approximate foreground removal 
has been performed for the standard or the \tszqe{} QE.

\subsubsection{Mock kinematic SZ signal}
\label{sec_mock_ksz}
We start by injecting a mock kSZ signal into the simulated CMB maps which is simply a temperature increment \mbox{$T = T_{\rm CMB} + T_{\rm kSZ}$} in the pixels that are within $2^{\prime}$ from the cluster center. 
The simulated map is then beam-convolved and processed as in the fiducial case. 
If unaccounted for, the kSZ signal, which is present in both the maps ($G$ and $L$), will introduce undesired correlations that are otherwise absent. 
The \iqe{} approach with $R1 = \innerradius$ 
should encompass this mock kSZ signal and must remain insensitive to it. 

To ensure that the LPF step before inpainting does not introduce filtering artifacts, we remove an estimated cluster kSZ signal using the aperture photometry (AP) technique \citep{planckxiii_14, alonso16}. 
In this technique, an estimate of $\hat{T}_{\rm kSZ}$ is obtained by computing the difference between the integrated temperature values in two concentric shells $T_{\theta_{1}} - T_{\theta_{2}}$  around the cluster where we set $\theta_{1} = 1.^{\prime}4$ and $\theta_{2} = \sqrt{2} \theta_{1}$. 
This kSZ estimate $\hat{T}_{\rm kSZ}$ is removed from the pixels within $\theta_{1}$ in the gradient map before the LPF step in the inpainting process.
We would like to emphasise that our goal here is simply to remove a significant portion of the kSZ signal for an efficient inpainting and not to reproduce the true kSZ profiles. 
As a result, the AP technique serves our purpose well here. 

The results of this test are presented in the top panel of Fig.~\ref{fig_mass_constraints_SZ_biases}. 
We use $T_{\rm kSZ} = 10 \mu K$.
No noticeable bias is detected for the \iqe{} while the standard and \tszqe{} are biased low by more than 17\%. 
The bias reduces to $\sim 6\%$ when a lower $\ell_{G} = 2000$ is adopted.
The low mass bias in the standard and \tszqe{} QE grow with the level of $T_{\rm kSZ}$ and is $>50\%$ for $T_{\rm kSZ} = 20 \mu K$. 
With the \iqe{} QE, the shift in the reconstructed lensing mass is smaller than the statistical uncertainty ($<0.4\sigma$) for $T_{\rm kSZ} = 20 \mu K$. 
While it is not realistic that all clusters in the sample have such large velocities and consequently a strong kSZ signal, this test serves to emphasise the robustness of the estimator. 


\subsubsection{Kinematic SZ from Sehgal simulations}
\label{sec_sehgal_kSZ}
In the previous section, we simply modified the temperature of the pixels within $2^{\prime}$ from the cluster center to input the kSZ signal into the simulated CMB maps. 
However, it is plausible that the true kSZ profile of the clusters and the correlated haloes has a much wider spread that can extend beyond the inpainting radius $R1$. 
If this is true, then the kSZ signal outside $R1$ will not only induce kSZ-bias in the lensing reconstruction but also contaminate the inpainting process.
To test this case, we use the kSZ signal corresponding to clusters in our sample from the \citetalias{sehgal10} simulations. 
As above, we use the AP technique to remove $\hat{T}_{\rm kSZ}$  in the gradient map before inpainting.
We also repeated this test with an amplified kSZ signal corresponding to clusters that are twice as massive ($\mvir = 4 \munits$). 
These two cases are shown in the second and the third panels of Fig.~\ref{fig_mass_constraints_SZ_biases}. 

Like in the case of mock kSZ, the lensing mass from the \iqe{} remains insensitive to the kSZ signal and the obtained lensing masses are $\le 0.1\sigma$ and $0.5\sigma$ from the true value in the two cases. 
The lensing mass from the standard and \tszqe{} QE are biased low by $\sim 10\%$ in the second panel (3\% for $\ell_{G} = 2000$ instead of $2700$) and the bias increases to $\gtrapprox 15\%$ (third panel) when the kSZ signal is amplified.  

When using the kSZ signal from even more massive clusters \mbox{$\mvir \ge 5.9 \munits$}, the lensing mass from the \iqe{} is low by 7.2\% ($1.3\sigma$). 
However, the number of haloes above this mass drops to less than 50 in the \citetalias{sehgal10} simulations and our results could be limited by the sample variance. 
Nevertheless, the bias becomes much smaller than the statistical error when a larger $R1 = 6^{\prime}$ is adopted for the inpainting process.

\subsubsection{Thermal SZ from Sehgal simulations}
\label{sec_sehgal_tSZ}
In the same spirit, we also checked the effect of tSZ signal from the cluster and correlated haloes \citep{vikram17, hill18} that extend outside the inpainting radius $R1$ using \citetalias{sehgal10} simulations. 
As in the kSZ case, we remove a  tSZ estimate from the gradient map before the LPF step in the inpainting process. 
Given that tSZ is much brighter and more extended than kSZ, we do not rely on a simple AP technique to filter tSZ. 
Instead, motivated by the work of \citet{patil19}, we fit a 2D circular Gaussian model to the cluster tSZ signal and subtract that from the CMB map before the LPF step. 
The Gaussian model has four free parameters: amplitude, $x$ and $y$ centroids, and the width. 
Note that we perform this filtering in the gradient $G$ map unlike \citet{patil19} in the $L$ map. 
While a Gaussian template may not represent the true profile of the cluster tSZ signal, the main objective of the template subtraction is to remove a significant amount of tSZ signal to reduce filtering artifacts from the LPF. 
We perform the fitting with the 150 GHz tSZ map that is devoid of CMB by combining the the 90 and 150 GHz channels. 

The tSZ tests are presented in the bottom panel of Fig.~\ref{fig_mass_constraints_SZ_biases}.
The \sqe{}, as expected, is heavily biased in the presence of tSZ. 
Reducing $\ell_{G} = 2000$ does not help much in the case of tSZ bias for the \sqe.
On the other hand, the \iqe{} estimator after removing a Gaussian tSZ model in the gradient map, works well even in the presence of tSZ. 
We obtain a lensing mass of \mbox{$\mrecov = 1.94 \pm 0.16 \munits$} consistent with the true cluster mass.
If we replace the Gaussian fitting with the AP technique like in the kSZ case, the recovered lensing mass is biased low by $3.5\sigma$ for our fiducial inpainting radius $R1 =\innerradius$. 
The bias is reduced when $R1$ is increased. 
As a further test, we amplified the tSZ signal by injecting signals corresponding to clusters with \mbox{$\mvir = 4 \munits$}. 
The \iqe{} returns a lensing mass of \mbox{$\mrecov = 2.18 \pm 0.29 \munits$} for this case.

For both the above masses, we note that the mass constraints are weakened (8.2\% as opposed to 5.6\% in Fig.~\ref{fig_mass_constraints}) when Sehgal tSZ haloes are added to the simulations. 
As observed previously in \citetalias{raghunathan18}, this extra variance is due to the presence of the cluster tSZ signal and adjacent haloes (both correlated and uncorrelated haloes in the line-of-sight) in the second leg $L$ of the QE. 
The residuals from the fitting process also contribute to the excess variance but they are sub-dominant compared to the tSZ signals in the $L$ map.
The extra variance can be reduced by removing a matched filter estimate of the tSZ from the second leg for the \tszqe{} and \iqe{} QE as demonstrated by \citet{patil19}.

\subsubsection{Discussion and applicability to real data}
\label{sec_discussion}
In the previous sections, we have shown that removing the cluster SZ signals using the inpainting technique performs better than simple gradient scale cuts in the QE. 
This is because adopting a gradient scale cut only removes the foreground power from scales smaller than $\ell > \ell_{G}$ leaving the foregrounds intact on large-scales $\ell \le \ell_{G}$. 
Since the foregrounds have non-zero off-diagonal elements in the harmonic space, unlike CMB, the foreground-induced correlations between the large-scales ($\ell \le \ell_{G}$) in the gradient $G$ map and the small-scales in the $L$ map remain unaltered. 
The level of bias from the residual foregrounds after the gradient cut depends on the foreground power relative to the experimental noise. 
For example, \citet{baxter18} adopted $\ell_{G} = 1500$ to reduce tSZ-induced lensing bias to $\sim 10\%$ for SPT-SZ experiment ($\Delta_{T} = 18$ \ukam) while for SPTpol, which is $4\times$ deeper than SPT-SZ, the tSZ-bias with $\ell_{G} = 1500$ is $\sim 50\%$ (see Fig. 2 of \citepalias{raghunathan18}). 
The bias from kSZ was much smaller than the statistical uncertainties in the above SPT analyses while we have shown in this work that it is not true for CMB-S4 datasets. 
This is consistent with the reduction in level of bias for $\ell_{G} = 2000$ vs $\ell_{G} = 2700$ in the \sqe: 6\% vs 17\% for mock kSZ (\S\ref{sec_mock_ksz}) and 3\% vs 10\% for  kSZ from \citetalias{sehgal10} simulations (\S\ref{sec_sehgal_kSZ}).

By contrary, inpainting completely removes the spatially-localized foreground signals present inside the inpainting radius. 
The foreground signals from haloes correlated to the cluster outside the inpainting radius can also be important but are highly sub-dominant. 
The mock kSZ case where the foreground signal is confined to the central $2^{\prime}$ region clearly illustrates this. 
It also indicates that the robustness of the \iqe{} QE does not come from the approximate subtraction of SZ signals outside the $R1$ in \iqe{} estimator. 

The cluster foreground signals used for the tests are extracted from N-body simulations \citepalias{sehgal10} that mimics the actual data expected from CMB experiments.
This ensures that we capture all of the real world effects like, for example, asymmetric cluster SZ profiles, SZ signals from correlated structures outside the cluster radius, offsets between the true cluster center and the SZ centers, scatter in the cluster SZ flux-mass relation, and varied range of cluster velocities and optical depth relevant for the cluster kSZ signals.
The inpainting radius $R1 = \innerradius$ used in this work was chosen conservatively.  
It encompasses the cluster foreground signals expected from clusters from the next generation CMB surveys \citep{benson14, SO18, S4_DSR_19}. 
The sizes of these clusters span roughly few arcminutes on the sky.
However, we have also shown that $R1$ must be increased slightly when dealing with extremely massive clusters. 
In general, we find $R1 \sim 2 \theta_{200c}$ to be a good choice for all clusters.


\section{Conclusion}
\label{sec_conclusion}
In this work we have presented a modification to the standard lensing QE by inpainting the large-scale CMB gradient at the cluster location. 
Using this estimator, we predict that the stacked mass of a cluster sample containing \howmanyclusters{} clusters (\mbox{$\mvir = \massval$} at $z = 0.7$) can be constrained to $\Delta M/M = 6.5\%$ with a CMB-S4 like low-noise ($\Delta_{T}$ = 1.5 \ukam) dataset without performing any foreground cleaning. 
We compared the estimator to the standard and \tszqe{} QE to show that the noise from inpainting is negligible and does not affect the lensing \snr. 
We studied the effect of cluster SZ biases on all the three estimators using realistic SZ signals from \citetalias{sehgal10} simulations. 
Our results demonstrate that the \iqe{} QE is robust against both cluster tSZ and kSZ signals and a similar result cannot be obtained using simple gradient scale cuts in the \sqe{} or the \tszqe{} QE.
In addition to the SZ-induced lensing biases, we have studied the effect of low mass bias arising from the underestimation of the CMB gradient due to the cluster convergence signal. 
We find that it is safe to use gradient modes up to $\ell \le \ell_{G}\ (= 2700)$ without worrying about the \ccbnoitalics{} when reconstructing the convergence signal, of clusters with \mbox{$\mvir \lessapprox 4\times10^{14}~\msol$}, using the future CMB surveys.
By doing so, we achieve a 14\% improvement in the mass constraints compared to the fiducial analysis with $\ell_{G} = 2000$. 

The \iqe{} estimator presented here will be viable to remove SZ biases for the small-scale cluster lensing reconstruction with the CMB temperature data from the near-term and future CMB experiments. 
The study about the effect of \ccbnoitalics{} performed in this work is also relevant for lensing reconstruction using the polarization data with the future datasets. 
In addition to the \sqe, the inpainting method can also be applied to the \tszqe{} and gradient inversion technique \citep{horowitz19} for an effective suppression of the kSZ-induced lensing biases. 

While not explicitly shown here, the \iqe{} QE can also be employed to handle the biases due to dust and synchrotron emissions from member galaxies present inside the cluster. 
But, we leave a careful consideration of these effects to a future work.
In addition to the biases considered here, there are other potential systematics that can produce dipole signals just like the cluster lensing.   
These are due to (a) the transverse motion of the cluster and correlated haloes referred commonly as the moving-lens effect \citep{lewis06, hotinli18, yasini19}; and (b) the rotational kSZ because of the rotation of the galaxy cluster \citep{lewis06, baxter15, baldi18, baxter19}. 
Both these effects may extend outside the inpainting radius.
However, in both these cases the direction of the dipole is oriented toward the direction of cluster motion or the rotation, while the CMB-cluster lensing dipole is oriented in the direction of the background CMB gradient. 
Since the background CMB gradient is not correlated to cluster motion or rotation, the moving-lens effect and the rotational kSZ should average down to zero in the stacked measurements presented here.

The inpainting technique is not limited to cluster lensing and, as demonstrated in an earlier work by \citet{benoitlevy13}, can also be used to fill the masked regions when reconstructing the CMB lensing power spectrum by the LSS. 
However, to deal with the SZ-induced biases, from unresolved haloes in particular, for the CMB lensing by LSS the methods prescribed by \citet{ferraro18, madhavacheril18, schaan18} or foreground bias-hardening techniques \citep{namikawa13, osborne14} will yield better results at the cost of a slightly higher lensing reconstruction noise.
The tSZ bias can be removed using a \tszqe{} gradient QE \citep{madhavacheril18}.
The bias from kSZ can be lowered by removing small-scale modes from lensing reconstruction in both legs \citep{ferraro18} or just the gradient leg with a degradation in the lensing \snr{}. 
Performing the lensing reconstruction using shear estimators is also another novel technique to remain insensitive to the foregrounds biases \citep{schaan18}.



\subsection*{Acknowledgements}
We are grateful to Nicholas Battaglia, Thomas Crawford, and Emmanuel Schaan for their constructive feedback on the manuscript. 
We also thank the anonymous referee for their useful suggestions that helped in shaping this manuscript better.

The UCLA authors acknowledge support from NSF grants AST-1716965 and CSSI-1835865. 
SP acknowledges support from Melbourne International Engagement Award (MIPP) and Laby Travel Bursary.
CR acknowledges the support from Australian Research Council's Discovery Projects scheme (DP150103208).
This work was performed in the context of the South Pole Telescope scientific program. SPT is supported by the National Science Foundation through grant PLR-1248097.  Partial support is also provided by the NSF Physics Frontier Center grant PHY-1125897 to the Kavli Institute of Cosmological Physics at the University of Chicago, the Kavli Foundation and the Gordon and Betty Moore Foundation grant GBMF 947. 
A part of this research was carried out at the Jet Propulsion Laboratory, California Institute of Technology, under contract with the National Aeronautics and Space Administration.
This research used resources of the National Energy Research Scientific Computing Center (NERSC), a DOE Office of Science User Facility supported by the Office of Science of the U.S. Department of Energy under Contract No. DE-AC02-05CH11231. 
We acknowledge the use of \texttt{HEALPix} \citep{gorski05} and \texttt{CAMB} \citep{lewis00} routines. 

\bibliographystyle{apsrev4-1}
\bibliography{spt_extract}

\end{document}